\documentclass{amsart}

\usepackage{amssymb,amsmath,amsthm,amsbsy}

\newtheorem{theorem}{Theorem}[section]
\newtheorem{lemma}[theorem]{Lemma}

\newtheorem{proposition}[theorem]{Proposition}
\newtheorem{claim}[theorem]{Claim}

\newenvironment{Proof}{\removelastskip\par\medskip
\noindent{\em Proof.} \rm}{\penalty-20\null\hfill$\square$\par\medbreak}

\numberwithin{equation}{section}

\newcommand{\Tr}{\text{Tr\,}}
\newcommand{\re}{\text{Re\,}}
\newcommand{\im}{\text{Im\,}}

\begin{document}
\setcounter{page}{1}

\title[]
{Universality of the local spacing distribution in certain ensembles
  of Hermitian Wigner matrices}
\author[K.~Johansson]{Kurt Johansson}

\address{
Department of Mathematics,
Royal Institute of Technology,
S-100 44 Stockholm, Sweden}

\email{kurtj@math.kth.se}

\begin{abstract}
Consider an $N\times N$ hermitian random matrix with independent
entries, not necessarily Gaussian, a so called Wigner matrix. It has
been conjectured that the local spacing distribution, i.e. the
distribution of the distance between nearest neighbour eigenvalues in
some part of the spectrum is, in the limit as $N\to\infty$, the same
as that of hermitian random matrices from GUE. We prove this conjecure
for a certain subclass of hermitian Wigner matrices.
\end{abstract}

\maketitle

\section{Introduction and main results}
Consider a probability measure $\mathbb P_N$ on the space of all $N\times
N$ hermitian matrices. We will be interested in the statistical
properties of the spectrum as $N$ becomes large, in particular in
features that are insensitive to the details of the particular 
sequence of probability measures we are considering. It is believed,
on the basis of numerical simulations, that for many types of
hermitian random matrix ensembles, i.e. choices of $\mathbb P_N$, the
local statistical properties of the eigenvalues are the same as for
the Gaussian Unitary Ensemble (GUE), where $d\mathbb P_N(M)=Z_N^{-1}
\exp(-\frac N2\Tr M^2) dM$. Here $dM$ is Lebesgue measure on the space
$\mathcal{H}_N\sim \mathbb R^{N^2}$ of all $N\times N$ hermitian matrices. The
asymptotic eigenvalue density as $N\to\infty$ (density of states) is
given by the Wigner semicircle law $\rho(t)=\frac
1{2\pi}\sqrt{(4-t^2)_+}$. Let $\rho_N(x_1,\dots,x_N)$ be the induced
probability density on the eigenvalues. The semicircle law is the limit  
of the one-dimensional marginal density as $N\to\infty$. The {\it m -
  point correlation function} 
\begin{equation}\label{1.1}
R_m^{(N)}(x_1,\dots,x_m)=\frac{N!}{(N-m)!}\int_{\mathbb
  R^{N-m}}\rho_N(x)dx_{m+1}\dots dx_N,
\end{equation}
is given by, \cite{Me} ch. 5, \cite{TW},
\begin{equation}\label{1.2}
R_m^{(N)}(x_1,\dots,x_m)=\det (K_N(x_i,x_j))_{i,j=1}^m,
\end{equation}
where the kernel $K_N(x,y)$ is given by
\begin{equation}\label{1.3}
K_N(x,y)=\frac{\kappa_{N-1}}{\kappa_N}\frac{p_N(x)p_{N-1}(y) -
  p_{N-1}(x)p_N(y)}{x-y}e^{-N(x^2+y^2)/4}.
\end{equation}
Here $p_N(x)=\kappa_Nx^N+\dots$ are the normalized orthogonal polynomials with
respect to the weight function $\exp(-Nx^2/2)$ on $\mathbb R$ (rescaled
Hermite polynomials). From these formulas, and Plancherel-Rotach 
asymptotics for the Hermite
polynomials it follows that 
\begin{equation}\label{1.4}
\lim_{N\to\infty}\frac{1}{(N\rho(u))^m}R_m^{(N)}(u+\frac{t_1}{N\rho(u)},\dots, 
u+\frac{t_m}{N\rho(u)})=\det\left(\frac{\sin\pi(t_i-t_j)}{\pi(t_i-t_j)}
\right) _{i,j=1}^m
\end{equation}
if $\rho(u)>0$. It has been proved, \cite{PS}, \cite{DKMVZ},
\cite{BI}, that this is also true in other invariant ensembles of the
form $d\mathbb P_N(M)=Z_N^{-1}\exp(-N\Tr V(M))dM$. The orthogonal
polynomials in (1.3) are then replaced by polynomials orthogonal with
respect to $\exp(-NV(x))$ on $\mathbb R$. That the ensemble is invariant
means that the probability measure is invariant under the conjugation 
$M\to U^{-1}MU$, with a unitary matrix $U$. Sufficient control of the
limit (\ref{1.4}) for all $m\ge 1$, makes it possible to determine the
asymptotic spacing distribution, i.e. distances between nearest
neighbour eigenvalues, see \cite{DKMVZ}. More precisely, let $\{t_N\}$ be a
sequence such that $t_N\to\infty$ but $t_N/N\to 0$ as $N\to\infty$ and
define, \cite{KS}, \cite{DKMVZ},
 $S_N(s,x)$, $s\ge 0$, $x\in\mathbb R^N$, to be the
symmetric function, which for $x_1<\dots<x_N$ is defined by
\begin{equation}\label{1.5}
S_N(s,x)=\frac 1{2t_N}\#\{1\le j\le
N-1\,;\,x_{j+1}-x_j\le\frac{s}{N\rho(u)},
|x_j-u|\le\frac{t_N}{N\rho(u)}\}.
\end{equation}
Given an hermitian matrix $M$ let $x_1(M)<\dots x_N(M)$ be its
eigenvalues; we write $x(M)=(x_1(M),\dots, x_N(M))$. Then it is proved
in \cite{DKMVZ} that
\begin{equation}\label{1.6}
\lim_{N\to\infty}\mathbb E_N[S_N(s, x(M))]=\int_0^s p(\sigma)d\sigma,
\end{equation}
for a large class of invariant ensembles. Here $p(\sigma)$ is the
density of the $\beta=2$ local spacing distribution, the {\it Gaudin
  distribution}, given by the probability density
\begin{equation}\label{1.7}
p(s)=\frac{d^2}{ds^2}\det(I-K)_{L^2(0,s)},
\end{equation}
where $K$ is the operator on $L^2(0,s)$ with kernel
$K(t,s)=\sin\pi(t-s)/\pi(t-s)$, the {\it sine kernel}, see \cite{Me}.

The aim of the present paper is to extend (\ref{1.4}) and (\ref{1.6}) to
other, non-invariant ensembles. It is conjectured, see \cite{Me} p.9,
that (\ref{1.4}) and (\ref{1.6}) should hold also 
for so called {\it Wigner matrices}
where the elements 
are independent but not necessarily Gaussian variables.
In this case the probability measure is
not invariant under conjugation by unitary matrices. For other
results on Wigner matrices see for example \cite{Ba}, \cite{Kh}, \cite{KKP},
\cite{Po}, \cite{SS1} and \cite{SS2}.
In particular, in \cite{So} the universality of the fluctuations of
the largest eigenvalue is established. To be more precise, consider
the complex random variables $w_{jk}$, $1\le j\le k$ with independent
laws $P_{jk}=P_{jk}^R\otimes P_{jk}^I$, where $P_{jj}^I=\delta_0$. 
Let $\mathcal{W}^p$, a class of Wigner ensembles,  
denote the class of all $\{P_{jk}\}_{1\le j\le k}$ which
satisfy 
\begin{equation}\label{1.8}
\int zdP_{jk}(z)=0\quad,\quad\int |z|^2dP_{jk}(z)=\sigma^2
\end{equation}
for all $1\le j\le k$, and furthermore
\begin{equation}\label{1.9}
\sup_{j,k}\int|z|^pdp_{jk}(z)<\infty.
\end{equation}
If $w_{kj}=\bar w_{jk}$, 
$W=(w_{jk})_{j,k=1}^N$ is an $N\times N$ {\it hermitian Wigner
  matrix }.  

Fix $a>0$ and let $\phi_a(t)=(\pi a^2)^{-1/2}\exp(-t^2/a^2)$ be a
Gaussian density function. Define $Q_{jk}^{R,I}=\phi_a\ast
P_{jk}^{R,I}$,
$1\le j<k$, $Q_{jj}^R=\phi_{a\sqrt{2}}\ast P_{jj}^R$, $j\ge 1$ and
$Q_{jj}^I=\delta_0$. Then $Q$ is also a Wigner ensemble and we let 
$\mathcal{W}^p_a$ 
denote the subclass of $\mathcal{W}^p$ obtained in this way. Note that
although $\mathcal{W}^p_a$ does not contain all Wigner ensembles it does
contain cases where the distribution of the matrix elements have very
different shapes, so in this sense it is rather broad,
and proving universality in $\mathcal{W}^p_a$ clearly shows that the
universality is not restricted to the invariant ensembles. Another way to
describe this ensemble of random matrices is as follows. Let $V$ be a
GUE-matrix with the probability measure $Z_N^{-1}\exp(-\frac 12\Tr
V^2)dV$ and let $W$ be an $N\times N$ Wigner matrix with distribution 
$P\in\mathcal{W}^p$, i.e. the law of $w_{jk}$ is $P_{jk}$. We will
assume that the variance $\sigma^2=1/4$, which can always be achieved
by rescaling.
Then $W+aV$  has the distribution $Q$,  and we write
\begin{equation}
M=\frac{1}{\sqrt{N}}(W+aV)\notag.
\end{equation}
We can think of this in terms of Dyson's Brownian motion model,
\cite{Dy}, $W+aV$ is obtained from $W$ by letting the matrix elements
execute a Brownian motion for a time $a^2$, see sect. 2.
If $P\in\mathcal{W}^p$ and $W$ is is an $N\times N$ hermitian matrix we let
$P^{(N)}$
denote the distribution of $H=W/\sqrt{N}=(h_{jk})$, i.e.
\begin{equation}
dP^{(N)}(H)=\prod_{1\le j\le k\le N} dP_{jk}(\sqrt {N}h_{jk})\notag.
\end{equation}
The matrix $M$ has the distribution $Q^{(N)}$, which is given by
\begin{equation}\label{1.10}
dQ^{(N)}(M)=2^{-N/2}\left(\frac{N}{\pi
    a^2}\right)^{N^2/2}\left(\int_{\mathcal{H}_N}e^{-\frac{N}{2a^2}\Tr
    (M-H)^2}dP^{(N)}(H)\right)dM,
\end{equation}
and this is the measure we will study. The asymptotic distribution of
the eigenvalues $x_1,\dots,x_N$ of $M$ is the semicircle law
\begin{equation}\label{1.11}
\rho(u)=\frac{2}{\pi(1+4a^2)}\sqrt{(1+4a^2-u^2)_+}.
\end{equation}
The following proposition will be proved in sect. 2 using an
argument from \cite{BH1}, \cite{BH2}.

\begin{proposition}\label{eigenmeas}
The symmetrized eigenvalue measure on $\mathbb R^N$ induced by $Q^{(N)}$
has a density
\begin{equation}\label{1.12}
\rho_N(x)=\int_{\mathcal{H}_N}\rho_N(x;y(H))dP^{(N)}(H)
\end{equation}
where
\begin{equation}\label{1.12'}
\rho_N(x;y)=\left(\frac{N}{2\pi
    a^2}\right)^{N/2}\frac{\Delta_N(x)}{\Delta_N(y)}
    \det(e^{-\frac{N}{2a^2}(x_j-y_k)^2})_{j,k=1}^N
\end{equation}
and $\Delta_N(x)=\prod_{1\le i<j\le N}(x_i-x_j)$ is the Vandermonde
determinant. 
\end{proposition}
The main result of the present paper is that for Wigner ensembles from
$\mathcal{W}^p_a$ we can prove (\ref{1.4}) and (\ref{1.6}), and thus extend the
universality to a rather broad class of Wigner matrices.

\begin{theorem}\label{corr} Fix $a>0$ and assume that $|u|\le \sqrt{1/2+2a^2}$.
Let  $R_M^{(N)}(x_1,\dots,x_m)$ be the correlation functions, defined by
  (\ref{1.1}), of the eigenvalue measure $\rho_N$, (\ref{1.12}),
for $Q^{(N)}$, (\ref{1.10}). Let $f\in L_c^\infty(\mathbb R^m)$, the set of all
  $L^\infty$ functions on $\mathbb R^m$ with compact support, and set for
  $x\in\mathbb R^N$ 
\begin{equation}
(Sf)(x)=\sum_{i_1,\dots,i_m} ' f(x_{i_1},\dots,x_{i_m})\notag,
\end{equation}
where the sum is over all distinct indices from $\{1,\dots, N\}$. If
$Q\in\mathcal{W}_a^p$ with $p>2(m+2)$, then
\begin{align}\label{1.12''}
&\lim_{N\to\infty}\int_{\mathcal{H}_N}(Sf)(N\rho(u)(x_1(M)-u),\dots
N\rho(u)(x_N(M)-u))dQ^{(N)}(M)\\
&\lim_{N\to\infty}\int_{\mathbb R^m}f(t_1,\dots,t_m)\frac 1{(N\rho(u))^m}
R_m^{(N)}(u+\frac {t_1}{N\rho(u)},\dots,u+\frac {t_m}{N\rho(u)})d^mt\notag\\
&=\int_{\mathbb
  R^m}f(t_1,\dots,t_m)\det(\frac{\sin\pi(t_i-t_j)}{\pi(t_i-t_j)})_{i,j=1}^m 
d^mt.\notag
\end{align}
\end{theorem}

 The condition on $u$ is made just to simplify the
  saddle-point argument in sect. 3; the result should hold for any $u$
  with $\rho(u)>0$.

We can also prove that the spacing distribution is the same as for
GUE.

\begin{theorem}\label{spac} Fix any $a>0$ and 
assume that $Q\in\mathcal{W}_a^{6+\epsilon}$, $\epsilon>0$. Let
$S_N(s,x)$ be defined by (\ref{1.5}). Then, for any $s\ge 0$,
\begin{equation}\label{1.13}
\lim_{N\to\infty}\int_{\mathcal{H}_N}
S_N(s,x(M))dQ^{(N)}(M)=\int_0^sp(\sigma)d\sigma,
\end{equation}
where $p(s)$ is given by (\ref{1.7}).
\end{theorem}
The theorems will be proved in sect. 4 after the preparatory work in
sect. 2 and 3.
\section{The correlation functions}
We will start by proving Proposition 1.1 using the
Harish-Chandra/Itzykson-Zuber formula following \cite{BH1},
\cite{BH2}. 
After that
we will give a formula for the correlation functions of $\rho_N(x;y)$,
which is very close to the formula in \cite{BH3}, but our derivation
will be different. A central role will be played by non-intersecting
one-dimensional Brownian motions and we will use the formulas of
Karlin and McGregor. Also we will discuss the relation to Dyson's
Brownian motion model. This connection can be found in \cite{Gr} and
we will only give an outline.

\begin{Proof}
Let $F(x)$ be a continuous symmetric function on $\mathbb R^N$. By
Fubini's theorem
\begin{equation}\label{2.1}
\int_{\mathcal{H}_N} F(x(M))dQ^{(N)}(M)=c_N^{(1)}\int_{\mathcal{H}_N}\left(
\int_{\mathcal{H}_N}F(x(M))e^{-\frac N{2a^2}\Tr
  (M-H)^2}dM\right)dP^{(N)}(H)
\end{equation}
with $c_N^{(1)}=2^{-N/2}(N/\pi a^2)^{N^2/2}$. In the right hand side
of (\ref{2.1}) we make the substitution $M=U^{-1}RU$, with $U\in U(N)$
and $R\in\mathcal{H}_N$, and then integrate over $U(N)$. If we use Fubini's
theorem again, we obtain
\begin{equation}
c_N^{(1)}\int_{\mathcal{H}_N}\left(
\int_{\mathcal{H}_N}F(x(R))\left(\int_{U(N)}e^{-\frac N{2a^2}\Tr
  (U^{-1}RU-H)^2}dU\right)dR\right)dP^{(N)}(H).\notag
\end{equation}
Here we have also used the fact that $dM=dR$. The integral over $U(N)$
can now be evaluated using the Harish-Chandra/Itzykson-Zuber formula,
\cite{Ha},\cite{IZ}, see also \cite{Me} A.5. We obtain the integral
\begin{equation}
c_N^{(1)}c_N^{(2)}\int_{\mathcal{H}_N}\left(
\int_{\mathcal{H}_N}F(x)\frac
1{\Delta_N(x)\Delta_N(y)}\det(e^{-N(x_j-y_k)^2/2a^2})_{j,k=1}^NdR\right)
dP^{(N)}(H),\notag
\end{equation}
where $y_1,\dots,y_N$ are the eigenvalues of $H$ and
$c_N^{(2)}=(a^2/N)^{N(N-1)/2} \prod_{j=1}^Nj!$. The integrand in the
middle integral depends only on the eigenvalues $x$ of $R$ and hence
we can integrate out the other degrees of freedom in the standard way,
\cite{Me} ch. 3, and obtain, after using Fubini's theorem,
\begin{align}\label{2.2}
&\int_{\mathcal{H}_N} F(x(M))dQ^{(N)}(M)\notag\\&=
c_N^{(1)}c_N^{(2)}c_N^{(3)}\int_{\mathcal{H}_N}\left(
\int_{\mathbb R^N}F(x)
\frac
{\Delta_N(x)}{\Delta_N(y)}\det(e^{-N(x_j-y_k)^2/2a^2})_{j,k=1}^Nd^Nx\right) 
dP^{(N)}(H)
\end{align}
with $c_N^{(3)}=\pi^{N(N-1)/2}\prod_{j=1}^N(j!)^{-1}$. We see that 
$c_N^{(1)}c_N^{(2)}c_N^{(3)}=(N/2\pi a^2)^{N/2}$ and since (\ref{2.2}) holds
for arbitrary bounded, continuous, symmetric $F(x)$ we have proved that the
symmetrized eigenvalue measure is given by \ref{1.12}. This proves
proposition \ref{eigenmeas}.
\end{Proof}

Let $p_t(x,y)$ be the transition probability of a Markov process
$X(t)$ on $\mathbb R$ with continuous paths. Consider $N$ independent
copies of the process $(X_1(t),\dots,X_N(t))$ and assume that this is
a strong Markov process in $\mathbb R^N$. Suppose that the particles
start at positions $y_1<\dots<y_N$ at time 0. The probability density
that they are at positions $x_1<\dots<x_N$ at time $S$ given that their
paths have not intersected anytime during the time interval $[0,S]$
is, by a theorem of Karlin and McGregor, \cite{KM}, 
\begin{equation}
\det(P_S(y_j,x_k))_{j,k=1}^N.\notag
\end{equation}
Hence, the conditional probability density that the particles are at
positions $y_1<\dots<y_N$ at time 0, at positions $x_1<\dots<x_N$ at
time $S$, at positions $z_1<\dots<z_N$ at
time $S+T$, given that their paths have not intersected in the time
interval $[0,S+T]$ is
\begin{equation}\label{2.3}
q_{S,T}(x;y;z)\doteq \frac 1{\mathcal{Z}_N}\det(P_S(y_j,x_k))_{j,k=1}^N
\det(P_T(x_j,z_k))_{j,k=1}^N,
\end{equation}
where
\begin{equation}
\mathcal{Z}_N=\int_{x_1<\dots<x_N}\det(P_S(y_j,x_k))_{j,k=1}^N
\det(P_T(x_j,z_k))_{j,k=1}^Nd^Nx;\notag
\end{equation}
we assume that $\mathcal{Z}_N>0$. Note that the expression (\ref{2.3}) is a
symmetric function of $x_1,\dots,x_N$, so we can regard it as a
probability measure on $\mathbb R^N$. Our next lemma shows that we can
obtain $\rho_N(x;y)$ defined by
(\ref{1.12'}) as a limit of the measure in (\ref{2.3}).

\begin{lemma}\label{qst}
Let $z_j=j-1$, $1\le j\le N$ and let $p_t(x,y)=(2\pi
t)^{-1/2}\exp((x-y)^2/2t)$ be the transition probability for
Brownian motion. Then, for any $x\in\mathbb R^N$ and $y_1<\dots ,y_N$,
\begin{equation}\label{2.4}
\lim_{T\to\infty}q_{S,T}(x;y;z)=\frac 1{(2\pi
  S)^{N/2}}\frac{\Delta_N(x)}{\Delta_N(y)}
\det(e^{-(x_j-y_k)^2/2S})_{j,k=1}^N\doteq q_S(x;y).
\end{equation}
\end{lemma}
Note that $\rho_N(x;y)=q_{a^2/N}(x;y)$.
\begin{Proof}
Write
\begin{align}\label{2.5}
&\det(P_S(y_j,x_k))_{j,k=1}^N
\det(P_T(x_j,z_k))_{j,k=1}^N\notag\\&=
\frac 1{(2\pi)^N(TS)^{N/2}}\det(e^{-(x_j-y_k)^2/2S})_{j,k=1}^N
\prod_{j=1}^Ne^{-\frac{x_j^2+z_j^2}{2T}}\det(e^{x_jz_k/2T})_{j,k=1}^N.
\end{align}
Note that $\mathcal{Z}_N$ is the conditional probability density of going from
$y_1<\dots ,y_N$ to $z_1<\dots ,z_N$ without collosions, i.e.
\begin{align}\label{2.6}
\mathcal{Z}_N&=\det(p_{S+T}(y_j,z_k))_{j,k=1}^N\notag\\
&=\frac 1{(2\pi)^{N/2}(S+T)^{N/2}}\prod_{j=1}^Ne^{-\frac{y_j^2+z_j^2}
    {2(S+T)}} \det(e^{y_jz_k/2(S+T)})_{j,k=1}^N.
\end{align}
Now, since $z_j=j-1$, we have two Vandermonde determinants in
(\ref{2.5}) and (\ref{2.6}). If we evaluate these, take the quotient
between (\ref{2.5}) and (\ref{2.6}) and then take the limit $T\to\infty$,
we obtain the right hand side of (\ref{2.4}).
\end{Proof}

Proposition 1.1 and lemma 2.1 establish a link between
  the eigenvalue distribution of $M=(W+aV)/\sqrt{N}$ and the
  non-intersecting Brownian paths. If we set $S=a^2/N$, then the right
  hand side of (\ref{2.4}) and (\ref{1.12'}) are identical;
  $y_1<\dots<y_N$ are the eigenvalues of $H=W/\sqrt{N}$. This relation
  can also be seen in another way, which we will now outline. Let
  $X(t)= (x_{jk}(t))_{j,k=1}^N$ be an $N\times N$ Hermitian matrix,
  where $\re x_{jk}(t)$, $\im x_{jk}(t)$, $j\le k$ are independent
  Brownian motions with variance $(1+\delta_{jk})/2$. Assume that
  $X(0)=H$ is distributed according to $P^{(N)}$. Then the distribution
  of $X(a^2/N)$ is the same as that of $M=(W+aV)/\sqrt{N}$. Following
  Dyson, \cite{Dy}, see also \cite{PR}, it is possible to derive a
  stochastic differential equation for the eigenvalues
  $\lambda_1(t),\dots,\lambda_N(t)$ of $X(t)$,
\begin{equation}\label{2.7}
d\lambda_i=dB_i+\sum_{k\neq i}\frac 1{\lambda_i-\lambda_k} dt,
\end{equation}
where $B_i$ are independent standard Brownian motions on $\mathbb R$, and
with the intial conditions $\lambda_i(0)=y_i$, $1\le i\le N$. We can
also consider the problem of non-intersecting Brownian motions in a
different way than that of Karlin and McGregor. Namely, let
$K=\{x\in\mathbb R^N\,;\,x_1<\dots<x_N\}$ and consider Brownian motion in
$\mathbb R^N$ starting at $y\in K$ and conditioned to remain in $K$
forever. As proved in \cite{Gr}, see also \cite{HW}, \cite{Pi}, 
if $\lambda_i$ are the
components of the $N$-dimensional conditioned Brownian motion they
satisfy the stochastic differential equation (\ref{2.7}) with the same
initial conditions. This gives another way to obtain (\ref{1.12'})
without using the Harish-Chandra, Itzykson/Zuber formula. Actually, we
can turn the argument around and give a proof of this formula.

We turn now to the computation of the correlation functions of the
right hand side of (\ref{2.4}), but we start more generally with
(\ref{2.3}). 
This can be analyzed using the techniques of \cite{TW}, compare
the analysis of the Schur measure, \cite{Ok}, in \cite{Jo1}, and see
also \cite{Jo2}. For
completeness, let us outline the result we need from \cite{TW}. Let
$(\Omega,\mu)$ be a measure space. Assume that
$\phi_j,\psi_j\in L^2(\Omega,\mu)$, $1\le j\le N$, and $f\in
L^\infty(\Omega,\mu)$. Set
\begin{equation}
Z_N[f]=\frac 1{N!}\int_{\Omega_N}\det(\phi_j(x_k))_{j,k=1}^N
\det(\psi_j(x_k))_{j,k=1}^N\prod_{j=1}^Nf(x_j)d\mu(x_j)\notag
\end{equation}
and
\begin{equation}
A=\left(\int_\Omega\phi_j(x)\psi_j(x)d\mu(x)\right)_{j,k=1}^N.\notag
\end{equation}
\begin{proposition}\label{TW} (\cite{TW}) Assume that $Z_N[1]\neq 0$. Then $A$
  is invertible and we can define
\begin{equation}\label{2.8}
K_N(t,s)=\sum_{j,k=1}^N\psi_k(t)(A^{-1})_{kj}\phi_j(s).
\end{equation}
Then, for any $g\in L^\infty(\Omega,\mu)$, 
\begin{equation}\label{2.9}
\frac{Z_N[1+g]}{Z_N[1]}=\det(I+K_Ng)_{L^2(\Omega)}.
\end{equation}
If we define a density on $\Omega^N$ by
\begin{equation}\label{2.10}
u_N(x)=\frac 1{N!Z_N[1]}\det(\phi_j(x_k))_{j,k=1}^N
\det(\psi_j(x_k))_{j,k=1}^N,
\end{equation}
then it has the correlation functions
\begin{equation}\label{2.11}
\frac{N!}{(N-M)!}\int_{\Omega^{N-m}}u_N(x)dx_{m+1}\dots
dx_N=\det(K_N(x_i,x_j))_{i,j=1}^N.
\end{equation}
\end{proposition}
\begin{Proof}
We will indicate the main steps in the proof of (\ref{2.9}) of which
(\ref{2.11}) is a consequence, see \cite{TW}. Set
\begin{equation}
B=\left(\int_\Omega\phi_j(x)\psi_j(x)g(x)d\mu(x)\right)_{j,k=1}^N.\notag 
\end{equation}
Then, by the formula
\begin{equation}
Z_N[f]=\det\left(\int_\Omega\phi_j(x)\psi_j(x)f(x)d\mu(x)\right)_{j,k=1}^N,
\notag
\end{equation}
which goes all the way back to \cite{An}, and which is not difficult
to prove by expanding the determinants, we see that
$\det A=Z_N[1]\neq 0$, so $A$ is invertible and
\begin{equation}\label{2.12}
\frac{Z_N[1+g]}{Z_N[1]}=\frac{\det(A+B)}{\det A}=\det(I+A^{-1}B).
\end{equation}
Now,
\begin{equation}
(A^{-1}B)_{jk}=\int_\Omega\psi_k(x)
\left(\sum_{\ell=1}^N(A^{-1})_{j\ell}\phi_\ell(x)g(x)\right)d\mu(x),\notag
\end{equation}
and we define $T:\mathbb C^N\to L^2(\Omega,d\mu)$ and $S:L^2(\Omega,d\mu)\to
\mathbb C^N$ by the kernels $T(x,k)=\psi_k(x)$ and
$S(j,x)=\sum_{\ell=1}^N (A^{-1})_{j\ell}\phi_\ell(x_g(x)$. Then, by
(\ref{2.12}) and a determinant identity, 
\begin{align}
\frac{Z_N[1+g]}{Z_N[1]}&=\det(I+ST)_{\mathbb
C^N}=\det(I+TS)_{L^2(\Omega,d\mu)}\notag\\
&=\det(I+K_Ng)_{L^2(\Omega,d\mu)}\notag,
\end{align}
with $K_N$ given by (\ref{2.8}). Note that $K_Ng$, which means first
multiplication by $g$ and then application of the operator on
$L^2(\Omega,d\mu)$ with kernel $K_N$, is a finite rank operator.
\end{Proof}
Observe now that if we take $\Omega=\mathbb R$, $d\mu(x)=dx$,
$\phi_j(x)=P_T(x,z_j)$ and $\psi_j(x)=p_S(y_j,x)$, then (\ref{2.3}) is a
probability density of the form (\ref{2.10}) and we can apply the
proposition. Note that
\begin{equation}
(A)_{jk}=\int_{\mathbb R}p_T(x,z_j)p_S(y_k,x)dx=p_{S+T}(y_k,z_j).\notag
\end{equation}
The kernel which gives the correlation functions is
\begin{equation}
K_N^{S,T}(u,v)=\sum_{k=1}^Np_S(y_k,v)\left(\sum_{j=1}^N
  (A^{-1})_{jk}p_T(u,z_j)\right).\notag
\end{equation}
Let $A_k(v)$ be the matrix we obtain from $A$ by replacing column $k$
by $(p_T(v,z_1)\dots p_T(v,z_N))^T$. Then, by Kramers' rule,
\begin{equation}\label{2.13}
K_N^{S,T}(u,v)=\sum_{k=1}^Np_S(y_k,v)\frac{\det A_k(v)}{\det A}.
\end{equation}
This formula and proposition \ref{TW} is the basis for the next
proposition. The result is closely related to the result derived in
\cite{BH3} by different methods.

\begin{proposition}\label{prop2.4} The correlation functions for 
$q_S(x;y)$ defined
  by (\ref{2.4}) are given by
\begin{align}\label{2.14}
R_m^{N}(x_1,\dots,x_m;y)&\doteq\frac{N!}{(N-m)!}\int_{\mathbb R^{N-m}}
q_S(x;y)dx_{m+1}\dots dx_N\\
&=\det(K_N^S(x_i,x_j;y))_{i,j=1}^m,\notag
\end{align}
where
\begin{align}\label{2.15}
&K_N^S(u,v;y)=\frac{e^{(v^2-u^2)/2S}}{(v-u)S(2\pi
    i)^2}\int_{\gamma}dz\int_{\Gamma}dw(1-e^{(v-u)z/S})\\
&\times\frac 1z\left(w+z-v-S\sum_j\frac{y_j}{(w-y_j)(z-y_j)}\right)
e^{(w^2-2vw-z^2+2uz)/2S}.\notag
\end{align}
Here $\gamma$ is the union of the curves $t\to -t+i\omega$, $t\in\mathbb
R$ and $t\to t-i\omega$, $t\in\mathbb R$ with a fixed $\omega>0$, and 
$\Gamma:\mathbb R\ni t\to it$.
\end{proposition}
\begin{Proof}
We have to show that with $p_t(u,v)=(2\pi t)^{-1/2}\exp(-(u-v)^2/2t)$
and $z_j=j-1$ the limit of the right hand side of (\ref{2.13}) as
$T\to\infty$ can be written as (\ref{2.15}). The result then follows
from lemma \ref{qst}, proposition \ref{TW} and the dominated
convergence theorem. We see that
\begin{equation}\label{2.16}
\det A=\frac
1{(2\pi(S+T))^{N/2}}\prod_{j=1}^Ne^{-\frac{z_j^2+y_j^2}{2(S+T)}}
\prod_{1\le i<j\le N}(e^{\frac{y_j}{S+T}}-e^{\frac{y_i}{S+T}})
\end{equation}
by the formula for a Vandermonde determinant. Let $\Gamma^\ast_M$ be the
curve $t\to t+iM$, $t\in \mathbb R$, $M$ fixed. Then
\begin{equation}
p_T(z_j,v)=\frac 1{\sqrt{2\pi
    T}}e^{-\frac{z_j^2}{2(S+T)}-\frac{v^2}{2T}}
\frac 1{\sqrt{2\pi}}\int_{\Gamma^\ast_M}e^{-\frac{\tau^2}2+z_j(\frac vT
    +i\tau\sqrt{\frac{S}{2T(S+T)}})}d\tau\notag
\end{equation}
Hence,
\begin{align}
\det A_k(v)&=\frac 1{2\pi\sqrt{T}}\frac 1{(2\pi (S+T))^{(N-1)/2}}
\left(\prod_{j=1}^Ne^{-\frac{z_j^2}{2(S+T)}}\right)\left(
\prod_{j\neq k}e^{-\frac{y_j^2}{2(S+T)}}\right)\notag\\
&\times e^{-\frac{v^2}{2T}}\int_{\Gamma^\ast_M} 
e^{-\frac{\tau^2}2}\det\tilde{A}_k(v)d\tau,\notag
\end{align}
where $\tilde{A}_k(v)$ is the matrix we get from
$(\exp(\frac{z_jy_k}{S+T}))_{j,k=1}^N$ by replacing column $k$ by
$(\exp(z_j(\frac vT+i\tau\sqrt{\frac{S}{2T(S+T)}})))_{j=1}^N$. Since
$z_{j}=j-1$ we have a Vandermonde determinant and we obtain
\begin{align}\label{2.17}
\det A_k(v)&=\sqrt{\frac{S+T}T}\frac 1{(2\pi (S+T))^{N/2}}
\left(\prod_{j=1}^Ne^{-\frac{z_j^2}{2(S+T)}}\right)\left(
\prod_{j\neq k}e^{-\frac{y_j^2}{2(S+T)}}\right)\notag\\
&\times e^{-\frac{v^2}{2T}}\frac 1{\sqrt{2\i}}\int_{\Gamma^\ast_M} 
e^{-\frac{\tau^2}2}
\prod_{1\le i<j\le N}(e^{\frac{y_j}{S+T}}-e^{\frac{y_i}{S+T}})d\tau,
\end{align}
where $y_k$ should be replaced by $(S+T)(\frac
vT+i\tau\sqrt{\frac{S}{2T(S+T)}}) $. Take the quotient of (\ref{2.16})
and (\ref{2.17}) and let $T\to\infty$. This gives
\begin{equation}
\lim_{T\to\infty}\frac{\det A_k(v)}{\det A}=\frac 1{\sqrt{2\pi}}
\int_{\Gamma^\ast_M}
e^{-\frac{\tau^2}2}\prod_{j\neq k}\left(\frac{v+i\sqrt{S}\tau-y_j}
{y_k-y_j}\right)d\tau.\notag
\end{equation}
Choose $M$ so that $v-\sqrt{S}M=L$, where $L$ is given, and make the
change of variables $w=v+i\sqrt{S}\tau$. Then
\begin{equation}
\lim_{T\to\infty}\frac{\det A_k(v)}{\det A}=
\frac 1{i\sqrt{2\pi S}}
\int_{\Gamma_L}e^{\frac{(w-v)^2}{2S}}\prod_{j\neq k}\left(\frac{
w-y_j}{y_k-y_j}\right)dw,\notag
\end{equation}
where $\Gamma_L:t\to L+it$, $t\in\mathbb R$. Thus, using (\ref{2.13}),
\begin{equation}
K_N^S(u,v;y)=\frac 1{2\pi iS}\sum_{k=1}^Ne^{-(y_k-u)^2/2S}
\int_{\Gamma_L}e^{\frac{(w-v)^2}{2S}}\prod_{j\neq k}\left(\frac{
w-y_j}{y_k-y_j}\right)dw.\notag
\end{equation}
Let $\gamma$ be a curve surrounding $y_1,\dots,y_N$ and choose $L$ so
large that $\gamma$ and $\Gamma$ do not intersect. The residue theorem
gives
\begin{equation}
\frac 1{2\pi i}\int_\gamma\frac{e^{-(z-u)^2/2S}}{w-z}\prod_{j=1}^N
\frac{w-y_j}{z-y_j}dz =
\sum_{k=1}^Ne^{-(y_k-u)^2/2S}\prod_{j\neq k}\left(\frac{
w-y_j}{y_k-y_j}\right)\notag
\end{equation}
for all $w\in\Gamma_L$. Thus,
\begin{equation}\label{2.18}
K_N^S(u,v;y)=\frac {e^{\frac{v^2-u^2}{2S}}}{(2\pi i)^2S}\int_\gamma
dz\int_{\Gamma_L} dwe^{\frac 1{2S}(w^2-2vw-z^2+2uz)}\frac 1{w-z}
\prod_{j=1}^N\frac{w-y_j}{z-y_j}.
\end{equation}
In (\ref{2.18}) we make the change of variables $z\to bz$, $w\to bw$
with $b\in\mathbb R$ close to 1. This will modify the contours but we can
use Cauchy's theorem to deform back to $\gamma$ and $\Gamma_L$. Now,
take the derivative with respect to $b$ and then put $b=1$. This gives
the equation
\begin{align}
0&=K_N^S(u,v;y)+\frac {e^{\frac{v^2-u^2}{2S}}}{(2\pi i)^2S^2}\int_\gamma
dz\int_{\Gamma_L} dw\frac 1{w-z}e^{\frac
  1{2S}(w^2-2vw-z^2+2uz)}\notag\\
&\times\left[w^2-z^2+uz-vw+S\sum_{j=1}^N\left(\frac{w}{w-y_j}-
\frac{z}{z-y_j}\right)\right]\prod_{j=1}^N\frac{w-y_j}{z-y_j}\notag.
\end{align}
This can be written
\begin{align}
&\frac{\partial}{\partial u}((u-v)K_N^S(u,v;y))=
-\frac {e^{\frac{v^2-u^2}{2S}}}{(2\pi i)^2S^2}\int_\gamma
dz\int_{\Gamma_L} dw\notag\\&\left[w+z-v-S\sum_{j=1}^N
\frac{y_j}{(w-y_j)(z-y_j)}\right]
e^{(w^2-2vw-z^2)/2S}e^{uz/S}\prod_{j=1}^N\frac{w-y_j}{z-y_j},\notag
\end{align}
and integration of this formula gives (\ref{2.15}). In this last formula
we can choose $L$ arbitrarily and take $\gamma$ to be the curve in the
proposition by using Cauchy's formula,
This completes the proof.
\end{Proof}

We now take $S=a^2/N$ and set
\begin{equation}\label{2.19}
\mathcal{K}_N(u,v;y)=e^{\frac{N(u^2-v^2)}{2a^2}+\omega(u-v)}K_N^{a^2/N}
(u,v;y), 
\end{equation}
where $\omega$ is a constant that will be specified later. Note that
we can replace $K_N^{a^2/N}$ with $\mathcal{K}_N$ in (\ref{2.14}) without
changing the correlation functions, so we can just as well work with
$\mathcal{K}_N$. Set
\begin{align}
&f_N(z)=\frac 1{2a^2}(z^2-2uz)+\frac 1N\sum_{j=1}^N\log(z-y_j)\notag\\
&g_N(z,w)=\frac 1{a^2z}\left(w+z-u-\frac{a^2}{N}
\sum_{j=1}^N\frac{y_j}{(w-y_j)(z-y_j)}\right)\notag\\
&h(z,w)=\frac{e^{\omega(u-v)}}{N\rho(u)(v-u)}e^{\frac N{a^2}(u-v)w}(
  e^{N(u-v)w/a^2}-e^{N(u-v)(w-z)/a^2})\notag,
\end{align}
so that
\begin{equation}\label{2.20}
\mathcal{K}_N(u,v;y)=N\rho(u)\int_\gamma\frac{dz}{2\pi
  i}\int_\Gamma\frac{dw}{2\pi i} h(z,w)g_N(z,w)e^{N(f_N(w)-f_N(z))}.
\end{equation}
These are the formulas we will use in the asymptotic analysis.
A straightforward computation shows that
\begin{equation}\label{2.21}
g_N(z,w)=\frac 1zf_N'(z)+\frac{f_N'(z)-f_N'(w)}{z-w}.
\end{equation}

\section{Asymptotics}
The eigenvalues $y_1,\dots,y_N$ of the Wigner matrix $H$ converge to
the semicircle law
\begin{equation}\label{3.1}
\sigma(t)=\frac 2{\pi}\sqrt{1-t^2},\quad |t|\le 1.
\end{equation}
In order to be able to perform the saddle point analysis of (\ref{2.20})
we need uniform control of the convergence of $f_N(z)$ to its limit
\begin{equation}\label{3.0}
f(z)=\frac 1{2a^2}(z^2-2uz)+\int_{-1}^1\log (z-t)\sigma(t)dt.
\end{equation}
In order to show this we must start with some probability
estimates. Write $\Omega_{R,\eta}=\{z\in\mathbb C\,;\,|\re z|\le R,
\eta\le |\im z|\le R\}$.
\begin{lemma}\label{lem3.1} Let $F\in L^\infty(\mathbb R^N)$ be symmetric
  and let $\eta>0$ and $R>0$ be given. Assume that $P\in\mathcal{W}^p$,
  $p>4$ and $0<\xi<\min(\frac 12-\frac 2p,\frac 1{16})$. Then, there
  is a probability measure $\tilde{P}^{(N)}$ on $\mathcal{H}_N$ such that
\begin{equation}\label{3.1'}
\left|\int_{\mathcal{H}_N}F(x(H))dP^{(N)}(H)-
\int_{\mathcal{H}_N}F(x(H))d\tilde{P}^{(N)}(H)\right|\le N^{2-p(\frac
12-\xi)}||F||_\infty ,
\end{equation}
and 
\begin{equation}\label{3.1''}
\sup_{z\in\Omega_{R,\eta}}\left|\frac
  1N\Tr\log(z-H)-\int_{-1}^1\log(z-t)\sigma(t)dt\right|\le CN^{-\xi}
\end{equation}
a.s. with respect to $\tilde{P}^{(N)}$.
\end{lemma}
\begin{Proof}
Given $P\in\mathcal{W}^p$ we introduce a cut-off $L>0$ and define a new
probability measure $P_L\in\mathcal{W}^p$ by
\begin{equation}
dP_{L,jk}^{R,I}(t)=\frac 1{d_{L,jk}}\chi_{[-L,L]}(t)dP_{jk}^{R,I}(t)
\quad, 1\le j\le k\notag
\end{equation}
where $d_{L,jk}$ is a normalization constant. Note that $P_{L,jk}$ is
supported in $K=[-L,L]^2$. Set $d^{(N)}_L=\prod_{1\le j\le k\le
  N}d_{L,jk}$. Then,
\begin{align}\label{3.2}
\left|\int_{\mathcal{H}_N}F(x(H))dP^{(N)}(H)-
\int_{\mathcal{H}_N}F(x(H))dP_L^{(N)}(H)\right|&\le
||F||_\infty(1-d^{(N)}_L)(1+\frac
1{d^{(N)}_L})\notag\\
&\le\frac{CN^2}{L^p}||F||_\infty
\end{align}
for some constant $C$. The last estimate follows from
\begin{equation}\label{3.2''}
1-d^{(N)}_L=P[\text{some\,} |W_{jk}|\ge L]\le N^2\sup_{1\le j\le
  k}\frac{E[|W_{jk}|^p]}{L^p}\le \frac{CN^2}{L^p}
\end{equation}
by (\ref{1.9}).
Set $D_N=\Omega_{R,\eta}\cap\frac 1N\mathbb Z^2$ and note that $\# D_N\le
CN^2$ for some constant $C$ that only depend on $R,\eta$. For a given
function $f$ set
\begin{equation}
A_N(f;\delta)=\{H\in\mathcal{H}_N\,;\,|\frac 1N\Tr
(f(H))-\int_{-1}^1f(t)d \sigma(t)|\le\delta\},\notag
\end{equation}
where $\sigma(t)$ is the semicircle law (\ref{3.2}). Set
\begin{equation}\label{3.2'}
A_N(\delta)=\bigcap_{z\in D_N}A_N(f_z,\delta),
\end{equation}
where $f_z(t)=\log(z-t)$ (principal branch). To estimate the
probability of $A_N(\delta)$ under $P^{(N)}$ we will use a result of
Guionnet and Zeitouni, \cite{GZ}. Let
\begin{equation}
|f|_{\mathcal{L}}=\sup_{t,s\in\mathbb R}\frac{|f(x)-f(y)|}{|x-y|},\notag
\end{equation}
and $||f||_{\mathcal{L}}=||f||_\infty+|f|_{\mathcal{L}}$. Then, by
\cite{GZ}, corollary 1.6a), and the discussion before this corollary,
given $\epsilon>0$, there are positive constants $C_0(\epsilon)$,
$C_1$ and $C_2$ such that if we write
\begin{equation}\label{3.3}
\delta_1(N)=C_1L^2|f|_{\mathcal{L}}N^{-1}+C_2(\epsilon)||f||_{\mathcal{L}}
N^{-1/4+\epsilon},
\end{equation}
then
\begin{equation}\label{3.4}
P_L^{(N)}\left[|\frac 1N\Tr
f(H)-\int_{-1}^1f(t)\sigma(t)dt|\ge\delta\right]\le
4\exp\left[-\frac{C_2N^2}{L^4 |f|_{\mathcal{L}}}(\delta-\delta_1(N))^2\right]
\end{equation}
for any $\delta>\delta_1(N)$.
Since under $P_L^{(N)}$ all $|H_{jk}|\le\sqrt{2}(L/\sqrt{N})$, the
spectral radius is $\le 2L$. Thus, the left hand side of (\ref{3.4})
is unchanged if we replace $f=f_z$ with $f=f_z^L(t)$, where
$f_z^L(t)=\log(z-t)$ if $|t|\le 2L$, 
$f_z^L(t)=\log(z-2L)$ if $t>2L$ and $f_z^L(t)=\log(z+2L)$ if
$t<-2L$. Now, $f_z^L(t)$ is Lipschitz and there is a constant $C_3$,
independent of $L$,
such that $|f_z^L(t)|_{\mathcal{L}}\le C_3 $ and
$||f_z^L(t)||_{\mathcal{L}}\le C_3(1+\log L)$ for all
$z\in\Omega_{z,\eta}$. Take $L=L_N=N^{1/2-\xi}$ and $\epsilon=1/6$ in
(\ref{3.3}). Then $\delta_1(N)\le CN^{-2\xi}$ and if we choose
$\delta=N^\xi$ in (\ref{3.4}) we obtain
\begin{equation}\label{3.5}
P_L^{(N)}\left[|\frac 1N\Tr
f_z(H)-\int_{-1}^1f_z(t)\sigma(t)dt|\ge N^{-\xi}\right]\le
c_1\exp(-c_2N^{2\xi})
\end{equation}
for some positive constants $c_1, c_2$. If we use (\ref{3.5}) we see
that the probability of the complement of the event in (\ref{3.2'})
can be estimated as
\begin{equation}\label{3.6}
P_{L_N}^{(N)}[A_N(N^{-\xi})^c]\le CN^2e^{-c_2N^{2\xi}}.
\end{equation}
Set 
\begin{equation}
d\tilde{P}^{(N)}(H)=(P_{L_N}^{(N)}[A_N(N^{-\xi})])^{-1}\chi_{A_N(N^{-\xi})}
(H) dP_{L_N}^{(N)}.\notag
\end{equation}
Note that $N^2/L_N^p=N^{2-p(1/2-\xi)}$, so combining (\ref{3.2}),
(\ref{3.2'}) and (\ref{3.6}) we obtain the estimate (\ref{3.1'}). From
the definition of $A_N(\delta)$ we see that (\ref{3.1''}) holds for
$z\in D_N$, but then a straightforward approximation argument extends
it to all $z\in\Omega_{R,\eta}$. This completes the proof of lemma
3.1.
\end{Proof}

We now come to the central asymptotic result.
\begin{lemma}\label{lem3.2} Let $\Omega_{R,\eta}$ be as above, let
  $\xi\in(0,1/2]$ and let $K$ be a compact subset of $\mathbb R$. Also
  let $u_N$ be a sequence such that $u_N\to u$ as
  $N\to\infty$. Furthermore, let $Y_{R,\eta}$ be the set of all
  $y\in\mathbb R^N$ such that 
\begin{equation}\label{3.7}
\sup_{z\in\Omega_{R,\eta}}\left|\frac 1N\sum_{j=1}^N\log(z-y_j)-
  \int_{-1}^1\log(z-t)\sigma(t)dt\right|\le CN^{-\xi}
\end{equation}
for some constant $C$ and all $N\ge 1$, where $\sigma(t)$ is given by
(\ref{3.1}). Then, we can find $R_0>0$, $\eta_0>0$ and a constant $C$
such that for all $y\in Y_{R_0,\eta_0}$, $\tau\in K$,
$|u|\le\sqrt{1/2+2a^2}$ and $N\ge 1$,
\begin{equation}
\left|\frac{1}{N\rho(u)}\mathcal{K}_N(u_N,u_N+\frac{\tau}{N\rho(u)};y)-
  \frac{\sin\pi \tau}{\pi\tau}\right|\le C(|u-u_N|+N^{-\xi}),
\end{equation}
where $\rho(u)$ is given by (\ref{1.11}).
\end{lemma}
\begin{Proof} 
It follows from the formula (\ref{2.20}) that
\begin{equation}\label{3.9}
\frac{1}{N\rho(u)}\mathcal{K}_N(u_N,u_N+\frac{\tau}{N\rho(u)};y)
=N\int_\gamma\frac{dz}{2\pi i}\int_\Gamma\frac{dw}{2\pi i}
h(z,w)g_N(z,w)e^{N(f_N(w)-f_N(z))} ,
\end{equation}
where $g_N(z,w)$ is given by (\ref{2.21}),
\begin{equation}
f_N(z)=\frac 1{2a^2}(z^2-2u_Nz)+\frac 1N\sum_{j=1}^N\log(z-y_j)
\notag
\end{equation}
and
\begin{equation}
h(z,w)=\frac{e^{\omega_0\tau}}{\tau}\left(e^{-\tau w/a^2\rho(u)}-
  e^{-\tau(w-z)/a^2\rho(u)}\right)\notag
\end{equation}
We have taken $\omega=\omega_0/N\rho(u)$, where $\omega_0$ is given by
(\ref{3.16}) below. The integral in (\ref{3.9}) will be analyzed using a
saddle point argument. It follows from (\ref{3.7}) and Cauchy's
integral formula that there is a constant $C$ such that for all $N\ge
1$, $\tau\in K$, $y\in Y_{R/2,2\eta}$ and $|u|\le\sqrt{1/2+2a^2}$,
\begin{align}\label{3.10}
&|f_N'(z)-f'(z)|\le C(N^{-\xi}+|u-u_N|)\\
&|f_N''(z)-f''(z)|\le CN^{-\xi}.\notag
\end{align}
A computation shows that,
\begin{equation}
f'(z)=\frac 1{a^2}(z-u)+2(z-\sqrt{z^2-1}).\notag
\end{equation}
Set $S(w)=(w+1/w)/2$ with inverse $S^{-1}(z)=z+\sqrt{z^2-1}$, where
$\sqrt{z^2-1}=\sqrt{z-1}\sqrt{z+1}$ (principal argument). The function
$S$ maps
$\{|w|>1\}$ to $\mathbb C\setminus [-1,1]$ and $|w|=1$ is mapped to
$[-1,1]$. Note that
\begin{equation}
f'(S(w))=\frac w{2a^2}+(2+\frac 1{2a^2})\frac 1w-\frac u{a^2}.\notag
\end{equation}
Write $u=\sqrt{1+4a^2}\cos\theta_c$, where $\theta_c\in [0,\pi]$. Our
assumption on $u$ means that $|\cos\theta_c|\le 1/2$. Note that
$f'(S(w))=0$ has the solutions $w_c^{\pm}=\sqrt{1+4a^2}\exp(\pm
i\theta_c)$. Hence the critical points for $f$ are
$z_c^{\pm}=S(w_c^{\pm})$.

We will now define some contours that we will use. Pick $\delta>0$
(small), see below. Set, for some $\epsilon>0$ (small),
$\gamma_1^+(t)=S(\sqrt{1+4a^2}e^{i\delta}-t)$, $-\infty<t\le 0$,
$\gamma_2^+(t)=S(\sqrt{1+4a^2}e^{it})$, $\delta\le t\le\theta_c-\epsilon$,
$\gamma_3^+(t)=S(\sqrt{1+4a^2}e^{it})$, 
$\theta_c-\epsilon\le t\le\theta_c+\epsilon$,
$\gamma_4^+(t)=S(\sqrt{1+4a^2}e^{it})$, 
$\theta_c+\epsilon\le t\le\pi-\delta$ and 
$\gamma_5^+(t)=S(\sqrt{1+4a^2}e^{i(\pi-\delta)}-t)$, $0\le t<\infty$.
Also, set $\gamma_j^-(t)=\overline{\gamma_j^+(t)}$, $1\le j\le
5$. Then, we can take $\gamma=\sum_{j=1}^5(\gamma_j^+-\gamma_j^-)
=\gamma^+-\gamma^-$ in
(\ref{3.9}). Let $t_0\in (1/\sqrt{1+4a^2}, 1)$ be such that $\im
S(t_0w_c^+)=\eta$, and write $\alpha=\re S(t_0w_c^+)$. Set, 
for some $\epsilon>0$ (small), $\Gamma_1^+(t)=\alpha+it$, $0\le
t\le\eta$, $\Gamma_2^+(t)=S(tw_c^+)$, $t_0\le t\le 1-\epsilon$,
$\Gamma_3^+(t)=S(tw_c^+)$, $1-\epsilon\le t\le 1+\epsilon$ and
$\Gamma_4^+(t)=S(tw_c^+)$, $1+\epsilon\le t$. Also, set
$\Gamma_j^-(t)= \overline{\Gamma_j^+(t)}$, $1\le j\le 4$. We can then
take $\Gamma=\sum_{j=1}^4(\Gamma_j^+-\Gamma_j^-)=\Gamma^+-\Gamma^-$ in
(\ref{3.9}). Set
\begin{equation}\label{3.10'}
L_N^{bd}(\tau;y)=N\int_{\gamma^b_3}\frac{dz}{2\pi i}
\int_{\Gamma^d_3}\frac{dw}{2\pi i}h(z,w)g_N(z,w)e^{N(f_N(w)-f_N(z))},
\end{equation}
where $b,d\in\{+,-\}$ and write
$L_N=L_N^{++}-L_N^{+-}-L_N^{-+}+L_N^{--}$.
\begin{claim}\label{cl3.3}
We can choose $R_0>0$, $\eta_0>0$ and $\epsilon,\delta>0$, so that 
$\gamma_3^++\gamma_3^-+\Gamma_3^++\Gamma_3^-$  lies in a neighbourhood
of $z_c^{\pm}$ which is included in $\Omega_{R_0/2,2\eta_0}$ and for
all $N\ge
1$, $\tau\in K$, $y\in Y_{r/2,2\eta}$ and $|u|\le\sqrt{1/2+2a^2}$,
\begin{equation}\label{3.11}
\left|\frac{1}{N\rho(u)}\mathcal{K}_N(u_N,u_N+\frac{\tau}{N\rho(u)};y)-
  L_N(\tau;y)\right|\le Ce^{-cN}
\end{equation}
with $c>0$
\end{claim}
The claim will be proved below. We will now use the claim to finish
the proof of lemma \ref{lem3.2}. It follows from (\ref{3.10}) that
there are critical points $z_N^\pm=S(w_N^\pm)$ for $f_N(z)$ such that 
\begin{equation}\label{3.12}
|z_N^\pm-z_c^\pm|\le C(N^{-\xi}+|u-u_N|).
\end{equation}
We can deform $\gamma_3^\pm$ ($\Gamma_3^\pm$) into contours 
$\gamma_N^\pm$ ($\Gamma_N^\pm$) such that the endpoints are unchanged,
$\gamma_N^\pm(0)=\Gamma_N^\pm(0)=z_N^\pm$ and $\gamma_N^\pm$
($\Gamma_N^\pm$) have $C^1$-distance $\le C(N^{-\xi}+|u-u_N|)$ to 
$\gamma_3^\pm$
($\Gamma_3^\pm$). We can also asume that these contours are
chosen so that $\gamma_N^\pm(t)=S(w_N^\pm e^{\pm it})$ and 
$\Gamma_N^\pm(t)=S(w_N^\pm(1+t))$ for $|t|\ll\epsilon$. 

We can now proceed in the standard way with a local 
saddle point argument in (\ref{3.10'}) and
prove that there is a constant $C$ such that 
\begin{align}\label{3.12'}
&\left|L_N^{bd}(\tau;y)-h(z_N^b,z_N^d)g_N(z_N^b,z_N^d)
\frac{2\pi}{(2\pi i)^2}\frac{(\gamma_N^b)'(0)(\Gamma_N^d)'(0)
  e^{N(f_N(z_N^b)-f_N(z_N^d))} }{\sqrt{f_N''(z_N^b)(\gamma_N^b)'(0)^2}
\sqrt{-f_N''(z_N^d)(\Gamma_N^d)'(0)^2}}\right|\\&\le\frac{C}{\sqrt{N}}
\notag
\end{align}
for all $N\ge
1$, $\tau\in K$, $y\in Y_{R_0,\eta_0}$ and $|u|\le\sqrt{1/2+2a^2}$.
Note that $z_N^+=\overline{z_N^-}$ and 
$f_N(z_N^+)-f_N(z_N^-)$ is purely imaginary. Now,
$(\gamma_N^b)'(0)=biS'(w_N^b)$, $(\Gamma_N^b)'(0)=w_N^bS'(w_N^b)$ and
a computation shows that
\begin{equation}
f_N''(z_N^b)(\gamma_N^b)'(0)^2=-f_N''(z_N^b)(\Gamma_N^b)'(0)^2=
-f_N''(z_N^b)S'(w_N^b)^2(w_N^a)^2,\notag
\end{equation}
which has a positive real part by (\ref{3.10}) and the fact that 
$f''(z_c^b)S'(w_c^b)^2(w_c^a)^2$ has a positive real part. From
(\ref{2.21}) we see that $g_N(z_N^b, z_N^d)=0$ if $b\neq d$ and 
$g_N(z_N^b, z_N^b)=f_N''(z_N^b)$. It follows that
\begin{equation}
\frac{g_N(z_N^b, z_N^b)(\gamma_N^b)'(0)(\Gamma_N^b)'(0)}
{\sqrt{f_N''(z_N^b)(\gamma_N^b)'(0)^2}
\sqrt{-f_N''(z_N^b)(\Gamma_N^b)'(0)^2}}=-bi.\notag
\end{equation}
Also, from (\ref{3.12}) it follows that
$|h(z_N^b,z_N^b)-h(z_c^b,z_c^b)|\le C(N^{-\xi}+|u-u_N|)$, and thus
(\ref{3.12'}) yields
\begin{equation}\label{3.13}
|L_N^{bd}(\tau;y)|\le\frac{C}{\sqrt{N}}
\end{equation}
if $b\neq d$ and
\begin{equation}\label{3.14}
\left|L_N^{bb}(\tau;y)+\frac{bh(z_c^b,z_c^b)}{2\pi i}\right|\le 
C(N^{-\xi}+|u-u_N|).
\end{equation}
Combining (\ref{3.10'}), (\ref{3.13}) and (\ref{3.14}) we obtain
\begin{equation}\label{3.15}
\left|L_N(\tau;y)+\frac{h(z_c^+,z_c^+)-h(z_c^-,z_c^-)}{2\pi
    i}\right|\le
C(N^{-\xi}+|u-u_N|).
\end{equation}
Now,
\begin{equation}
h(z_c^\pm,z_c^\pm)=\frac{e^{\omega_0\tau}}{\tau}\left(e^{-\tau
    z_c^\pm/a^2\rho(u)}-1\right)\notag
\end{equation}
and a computation shows that
\begin{equation}\label{3.16}
\frac{z_c^\pm}{a^2\rho(u)}=\pi\frac{1+2a^2}{2a^2}\cot\theta_c\pm\pi
i\doteq\omega_0\pm \pi i.
\end{equation}
Thus (\ref{3.15}) becomes
\begin{equation}
\left|L_N(\tau;y)-\frac{\sin\pi \tau}{\pi\tau}\right|\le
C(N^{-\xi}+|u-u_N|).
\notag
\end{equation}
If we combine this estimate with (\ref{3.11}) we see that the lemma is
proved.
\end{Proof}
It remains to prove claim \ref{cl3.3}.
\begin{Proof}
Let $\gamma_\ast^\pm=\sum_{j\neq 3}\gamma_j^\pm$ and 
$\Gamma_\ast^\pm=\sum_{j\neq 3}\Gamma_j^\pm$. We have to estimate
\begin{equation}
I_1^{bd}=N\int_{\gamma_\ast^b}|dz|\int_{\Gamma^d}|dw||h(z,w)||g_N(z,w)|
e^{N\re(f_N(w)-f_N(z_c^d))-N\re(f_N(z)-f_N(z_c^b))},\notag
\end{equation}
and
\begin{equation}
I_2^{bd}=N\int_{\gamma^b}|dz|\int_{\Gamma_\ast^d}|dw||h(z,w)||g_N(z,w)|
e^{N\re(f_N(w)-f_N(z_c^d))-N\re(f_N(z)-f_N(z_c^b))},\notag
\end{equation}
where $b,d\in\{+,-\}$. Note that $f_N(z_c^+)-f_N(z_c-)$ is purely
imaginary. We will concentrate on $I_1^{++}$ since the other cases are
similar. 

Using the inequality
\begin{equation}
\left|\frac{w-y_j}{z-y_j}\right|=\left|1+\frac{w-z}{z-y_j}\right|\le
1+C(|w|+|z|)\notag
\end{equation}
it is not difficult to see that there are constants $C_1$ and $C_2$
such that
\begin{equation}\label{3.16'}
|h(z,w)||g_N(z,w)|e^{N\re (f_N(w)-f_N(z))}\le
C_1E^{C_2N(|z|+|w|)+N(\re(w^2-2uw)-\re(z^2-2uz))/2a^2}
\end{equation}
for all $y\in\mathbb R^N$, $\tau\in K$ and $|u|\le\sqrt{1/2+2a^2}$. Note
that $|\im z|\ge c>0$ for all $z\in\gamma$. (The constant $c$ depends
on the $\delta$ in the definition of $\gamma$, but as we will see
below $\delta$ depends only on the parameter $a$ in the problem.) From
the estimate (\ref{3.16'}) it follows that by picking $R=R_0$
sufficiently large, the contribution to $I_1^{++}$ from $z$ and/or $w$
outside $\Omega_{R_0,0}$ is $\le e^{-N}$. Thus we can assume that
$z,w\in \Omega_{R_0,0}$. Next, we will derive the other estimates we
will need to prove the claim. 

Assume that $z\in\Omega_{R_0,\eta}$ and $w\in\Gamma_1^+$. Then,
\begin{align}
&|g_N(z,w)e^{Nf_N(w)}|\notag\\
&\le C\left(1+\frac 1N\sum_{k=1}^N\frac
  1{|w-y_k|}\right)\prod_{j=1}^N|w-y_j| e^{N\re(w^2-2uw)/2a^2}\notag\\
&\le C\left(1+\frac 1N\sum_{j=1}^N\frac 1{|\alpha+i\eta-y_j|}\right)
\prod_{j=1}^N|\alpha+i\eta-y_j|e^{N\re(w^2-2uw)/2a^2}\notag\\
&\le Ce^{N[\re f_N(\alpha+i\eta)+\re(w^2-2uw)-((\alpha+i\eta)^2-2u
(\alpha+i\eta))]/2a^2}.\notag
\end{align}
If we use (\ref{3.7}) and the definition of $f_N$ we obtain
\begin{align}\label{3.17}
&|g_N(z,w)e^{N(f_N(w)-f_N(z_c^+))}|\notag\\
&\le
Ce^{cN(N^{-\xi}+|u-u_N|)+N\eta^2/2a^2+N\re(f(\alpha+i\eta)-f(z_c^+))/2a^2}
\end{align}
for $z\in\Omega_{R_0,\eta}$ and $w\in\Gamma_1^+$.

We will now compute how $\re f(z)$ changes along $\gamma$. Assume that
$\theta_c\ge 0$, the other case is analogous. Consider
$\gamma(\theta)= S(\sqrt{1+4a^2}e^{i\theta})$,
$\delta\le\theta\le\pi-\delta$. A computation, using the fact that
$f'(\gamma(\theta_c))=0$ gives $\re\frac
d{d\theta}f(\gamma(\theta))=\frac {1+2a^2}{2a^2}\sin\theta
(\cos\theta_c -\cos\theta)$. From this we see that there is a constant
$c_0>0$ such that
\begin{equation}\label{3.18}
\re (f(\sqrt{1+4a^2}e^{i\theta})-f(z_c^+))\ge c_0(\theta-\theta_c)^2.
\end{equation}
Next, consider  $\gamma_1(t)=S(\sqrt{1+4a^2}e^{i\delta}-t)$, $t\le
0$. If we write $\omega_\delta=\sqrt{1+4a^2}e^{i\delta}$, then
\begin{equation}
\frac {d}{dt}f(\gamma_1(t))=-\frac
1{4a^2}[\omega_\delta-t-2u+\frac{1+4a^2}{\omega_\delta-t}][1-\frac
1{(\omega_\delta-t)^2}] .
\notag
\end{equation}
Set $\omega_\delta-t=s(t)e^{i\theta(t)}$. A computation shows that
\begin{align}\label{3.19}
&\re\frac {d}{dt}f(\gamma_1(t))
=-\frac{1}{4a^2\sqrt{1+4a^2}}\left\{\left[(s(t)+\frac 1{s(t)})\cos
      \theta(t)-2\cos\theta_c\right]\right.
\\&\left.\times\left[1+4a^2-\frac 1{s(t)^2}\cos
      2\theta(t) \right]
-\frac 1{s(t)^2}\sin 2\theta(t)(s(t)-\frac
1{s(t)})\sin\theta(t)\right\}.\notag
\end{align}
Note that $\sin\theta(t)=s(t)^{-1}\sqrt{1+4a^2}\sin\delta$. It follows
that the right hand side of (\ref{3.19}) equals
\begin{align}\label{3.19'}
&-\frac{1}{4a^2\sqrt{1+4a^2}}\left\{\left[(s(t)+\frac 1{s(t)})
(1+4a^2-\frac 1{s(t)^2}+2\frac{(1+4a^2)\sin^2\delta}{s(t)^4})\right.\right.\\&\left.\left.- 
2\frac{(1+4a^2)\sin^2\delta}{s(t)^4}(s(t)-\frac
1{s(t)})\right]\cos\theta(t)
-2(1+4a^2-\frac
1{s(t)^2}+2\frac{(1+4a^2)\sin^2\delta}{s(t)^4})\cos\theta_c \right\}\notag
\end{align}
and this is
\begin{equation}
\le -\frac{1}{4a^2\sqrt{1+4a^2}}(1+4a^2-\frac 1{s(t)^2})
\left[(s(t)+\frac
  1{s(t)})\cos\theta(t)-2(1+\frac{1+4a^2}{2a^2}\sin^2\delta)
  \cos\theta_c\right],\notag
\end{equation}
since $s(t)\ge 1$. Choose $\delta\le \theta_c/4$ so that
\begin{equation}
(1+\frac{1+4a^2}{2a^2}\sin^2\delta)
  \cos\theta_c\le\cos\frac{\theta_c}2.\notag
\end{equation}
Since $s(t)+1/s(t)\ge 2$ and $\theta(t)\le\delta$ we see that there is
a constant $c_0>0$ such that
\begin{equation}\label{3.21}
\re\frac {d}{dt}f(\gamma_1(t))\le -c_0.
\end{equation}
For $\gamma_5(t)=S(\sqrt{1+4a^2}e^{i(\pi-\delta)}-t)$, $t\ge 0$, we
still have the formula (\ref{3.19'}) with $\gamma_1(t)$ replaced by 
$\gamma_5(t)$ and, since $\pi-\delta\le\theta(t)\le\pi$, we see that
the right hand side is
\begin{equation}
\ge\frac 1{\sqrt{1+4a^2}}[(s(t)+\frac
1{s(t)})\cos(\pi-\theta(t))+2\cos\theta_c]
\end{equation}
and consequently there is a constant $c_0>0$ such that
\begin{equation}\label{3.22}
\re\frac {d}{dt}f(\gamma_5(t))\ge c_0.
\end{equation}
Consider now how $\re f(w)$ changes along $\Gamma^+$. Set
$\Gamma(t)=S(tw_c^+)$, $t\ge t_0$. A computation gives
\begin{equation}
\re\frac
{d}{dt}f(S(tw_c))=\frac{1-t}{2a^2t^2}[1+t(1+4a^2)-(t^2(1+4a^2)+\frac
1t)\cos 2\theta_c].\notag
\end{equation}
Now, since $|u|\le\sqrt{1/2+2a^2}$, it follows that 
$\cos 2\theta_c\le 0$ and thus
\begin{align}\label{3.23}
&\re\frac
{d}{dt}f(S(tw_c))\ge \frac{1-t}{2a^2t^2}(1+t(1+4a^2))\quad\text{if
  $t_0\le t\le 1$}\notag\\
&\re\frac
{d}{dt}f(S(tw_c))\le \frac{1-t}{2a^2t^2}(1+t(1+4a^2))\quad\text{if
  $t\ge 1$}.
\end{align}
The first of these estimates can be used to show that if we pick
$\eta=\eta_0$ sufficiently small, then
\begin{equation}
\eta^2+\re (f(\alpha+i\eta)-f(z_c^+))\le -c_0\notag
\end{equation}
for some positive $c_0$. If we use this in (\ref{3.17}) we obtain
\begin{equation}\label{3.24}
|g_N(z,w)e^{N(f_N(w)-f_N(z_c^+))}|\le Ce^{-c_0'N}
\end{equation}
for some positive $c_0'$. We can now use 
(\ref{3.18}), (\ref{3.21}), (\ref{3.22}), (\ref{3.23}) and
(\ref{3.24}) to estimate $I_1^{++}$ and see that it is $\le Ce^{-cN}$
for some positive $c$.
\end{Proof}

\section{Proof of the theorems}
We start with the proof of theorem \ref{corr}. 
\begin{Proof}
By proposition 1.1 and Fubini's theorem the integral in the left hand side of
(\ref{1.12''}) can be written
\begin{equation}\label{4.1}
\int_{\mathcal{H}_N}\left(\int_{\mathbb R^N}\rho_N(x,y(H))(Sf)
    (N\rho(u)(x_1-u), \dots,N\rho(u)(x_N-u))d^Nx\right)dP^{(N)}(H)
\end{equation}
Note that $||S(f)||_\infty\le N^m||f||_\infty$. Since
$\rho_N(x,\cdot)$ is a probability density on $\mathbb R^N$ we can use
lemma 3.1 to replace the expression in (\ref{4.1}) by
\begin{equation}\label{4.2}
\int_{\mathcal{H}_N}\left(\int_{\mathbb R^N}\rho_N(x,y(H))(Sf)
    (N\rho(u)(x_1-u), \dots,N\rho(u)(x_N-u))d^Nx\right)d\tilde{P}^{(N)}(H)
\end{equation}
with an error $\le CN^m||f||_\infty N^{2-p(1/2-\xi)}=o(1)$, since
$p>2(m+2)$, provided we choose $\xi$ small enough. Now, since
$\rho_N(x,\cdot)$ is symmetric it follows from (\ref{1.12'}),
(\ref{2.4}), (\ref{2.14}) and (\ref{2.19}) that the expression
in (\ref{4.2}) can be written
\begin{align}\label{4.3}
\int_{\mathcal{H}_N}\int_{\mathbb R^m} &f(t_1,\dots,t_m)\\
&\times\det(\frac
    1{N\rho(u)}
    \mathcal{K}(u+\frac{t_i}{N\rho(u)},u+\frac{t_j}{N\rho(u)};y(H)))_{i,j=1}^m 
d^mtd\tilde{P}^{(N)}(H).\notag
\end{align}
Since $f$ has compact support and we know that (\ref{3.1''}) holds
a.s. $[\tilde{P}^{(N)}]$ it follows from lemma \ref{lem3.2}, with
$u_N=u+t_i/N\rho(u)$, $\tau=t_j-t_i$, that
\begin{equation}
\left|\mathcal{K}(u+\frac{t_i}{N\rho(u)},u+\frac{t_j}{N\rho(u)};y(H))
  -\frac{\sin\pi(t_i-t_j)}{\pi(t_i-t_j)}\right|\le CN^{-\xi},\notag
\end{equation}
for a.a. $[\tilde{P}^{(N)}]$ and all $(t_1,\dots,t_m)$ in the support
of $f$. Thus we can take the limit as $N\to\infty$ in (\ref{4.3}) and
obtain the right hand side of (\ref{1.12''}). This completes the proof.
\end{Proof}

Before proving theorem 1.3 we need some preliminary results on the
level spacing distribution. Let $\rho_N(x)$ be a symmetric probability
density on $\mathbb R^N$ with correlation functions defined by
(\ref{1.1}). Assume that $R_1^{(N)}/N\to\rho(t)$ (weakly) as
$N\to\infty$, so that $\rho(t)$ is the asymptotic density. Let $u$ be
a given point such that $\rho(u)>0$, and let $t_N$ be a sequence such
that $t_N\to\infty$ but $t_N/N\to 0$ as $N\to\infty$. Set, for $|r|\le
1/2$,
\begin{equation}
\mathcal{R}_m^{(N)}(\sigma_1,\dots,\sigma_m;r)=
\frac 1{(N\rho(u))^m}R_m^{(N)}(u+\frac{2t_Nr+\sigma_1}
{N\rho(u)},u+\frac{2t_Nr+\sigma_m}{N\rho(u)})\notag
\end{equation}
and let $\mathcal{R}_m(\sigma_1,\dots,\sigma_m)$ be the limiting
correlation functions, which we assume are continuous, symmetric and
translation invariant. Assume that, for each $s\ge 0$,
\begin{equation}\label{4.4}
D_N(s)=\sum_{m=N+1}^\infty\frac{s^m}{m!}\sup_{|\sigma_j|\le s}
|\mathcal{R}_m(\sigma_1,\dots,\sigma_m)|<\infty.
\end{equation}
Set 
\begin{equation}
H(s)=\sum_{m=0}^\infty\frac{(-1)^m}{m!}\int_{[0,s]^m}
\mathcal{R}_m(\sigma_1,\dots,\sigma_m)d^m\sigma\notag
\end{equation}
(the probability of no particle in $[0,s]$), which is well defined by
(\ref{4.4}). Also, set
\begin{equation}\label{4.5}
\epsilon_m^{(N)}=\sup_{|\sigma_j|\le s, |r|\le 1/2}
|\mathcal{R}_m^{(N)}(\sigma_1,\dots,\sigma_m;r)-
\mathcal{R}_m(\sigma_1,\dots,\sigma_m)|.
\end{equation}

\begin{proposition}\label{prop4.1}
Let $S_N(s,x)$ be defined by (\ref{1.5}). Then
\begin{equation}\label{4.6}
\left|\int_{\mathbb R^N}S_N(s,x)\rho_N(x)d^Nx-\int_0^s H''(u)du\right|\le
D_N(s) +\sum_{m=2}^N\frac{s^{m-1}}{(m-1)!}\epsilon_m^{(N)}.
\end{equation}
\end{proposition}
\begin{Proof}
We first show that
\begin{equation}\label{4.7}
\int_0^s H''(u)du=\sum_{m=2}^N\frac{s^{m-1}}{(m-1)!}\int_{[0,s]^{m-1}}
\mathcal{R}_m(0,\tau_2,\dots,\tau_m)d\tau_2\dots d\tau_m,
\end{equation}
see \cite{DKMVZ}. Since $\mathcal{R}_m$ is translation invariant and
symmetric by assumption, we have
\begin{align}\label{4.7'}
H'(u)&=\lim_{\epsilon\to 0}\frac 1{\epsilon}\sum_{m=0}^\infty
\frac{(-1)^m}{m!} \int_{[-\epsilon,u]^m\setminus [0,u]^m}
\mathcal{R}_m(x_1,\dots,x_m)d^mx\\
&=\lim_{\epsilon\to 0}\sum_{m=0}^\infty
\frac{(-1)^m}{m!}\frac 1{\epsilon}\left( m\int_{[-\epsilon,0]\times
    [0,u]^{m-1}}\mathcal{R}_m(x_1,\dots,x_m)d^mx\right)\notag \\
&=\sum_{m=1}^\infty\frac{(-1)^m}{(m-1)!}\int_{[0,u]^{m-1}}
\mathcal{R}_m(0, x_2,\dots,x_m)d^{m-1}x\notag ,
\end{align}
where we have also used (\ref{4.4}) and the continuity of
$\mathcal{R}_m$. Continuing in the same way we see that $H(u)$ is
actually a $C^\infty$ function, in particular $H''(u)$ is well defined
and continuous. From (\ref{4.7'}) we get
\begin{equation}
H'(s)=-\mathcal{R}_m(0)+\sum_{m=2}^\infty
\frac{(-1)^m}{(m-1)!}\int_{[0,s]^{m-1}}
\mathcal{R}_m(0, x_2,\dots,x_m)d^{m-1}x.\notag
\end{equation}
Hence $H'(0)=-\mathcal{R}_m(0)$ and we see that the right hand side of
(\ref{4.7}) equals $H'(u)-H'(0)$, which is what we wanted to prove.

It is proved in \cite{DKMVZ}, using a result from \cite{KS}, that
\begin{align}
&\int_{\mathbb R^N}S_N(s,x)\rho_N(x)d^Nx\notag\\&=
\sum_{m=2}^N\frac{(-1)^m}{(m-1)!}\int_{-1/2}^{1/2}dr\int_{[0,\min(s,
  (1-2r)t_N)]^{m-1}}
\mathcal{R}_m^{(N)}(0,\sigma_2,\dots,\sigma_m;r)d^{m-1}\sigma.\notag 
\end{align}
Hence, the estimate (\ref{4.6}) follows from  
(\ref{4.4}), (\ref{4.5}) and (\ref{4.7}).
\end{Proof}

We turn now to the proof of theorem \ref{spac}.
\begin{Proof}
Just as in the proof of theorem \ref{corr} above we see that 
since $P\in\mathcal{W}^{6+\epsilon}$ and $||S_N||_\infty\le N/2t_N$,
\begin{align}\label{4.8}
&\left|\int_{\mathcal{H}_N}S_N(s,x(M))dQ^{(N)}(M)-
\int_{\mathcal{H}_N}\left(\int_{\mathbb
    R^N}S_N(s,x)\rho_N(x;y(H))d^Nx\right) d\tilde{P}^{(N)}(H)\right|\\
&\le C\frac N{t_N}N^{2-(6+\epsilon)(1/2-\xi)}\le \frac{C}{t_N},\notag
\end{align}
if we take $\xi$ sufficiently small, and also that (\ref{3.1''})
holds. From proposition \ref{eigenmeas}, (\ref{2.4}) and proposition
\ref{prop2.4} we know the correlation functions of $\rho_N(x;y)$, and
if we take $u_N=u+(2t_Nr+\sigma_i)(N\rho(u))^{-1}$ in lemma
\ref{lem3.2} we see that
\begin{align}\label{4.9}
\left|\frac
    1{N\rho(u)}
    \mathcal{K}(u+\frac{2t_Nr+\sigma_i}{N\rho(u)},
u+\frac{2t_Nr+\sigma_j}{N\rho(u)};y(H))-
\frac{\sin\pi(\sigma_i-\sigma_j)}{\pi(\sigma_i-\sigma_j)} \right|&\le
    C(\frac{t_N}N +N^{-\xi})\\&\doteq \omega_N\notag
\end{align}
for a.a. $H$ $[\tilde{P}^{(N)}]$. Thus, the limiting correlation
functions are
\begin{equation}
\mathcal{R}_m(\sigma_1,\dots,\sigma_m)-
\det\left(\frac{\sin\pi(\sigma_i-\sigma_j)}
{\pi(\sigma_i-\sigma_j)} \right)_{i,j=1}^m.\notag
\end{equation}
Since the matrix in the determinant is positive definite it follows
from the Hadamard inequality that
\begin{equation}
D_N(s)\le \sum_{m=N+1}^\infty\frac{s^m}{m!}.\notag
\end{equation}
Also, since
\begin{equation}
\mathcal{R}_m^{(N)}(\sigma_1,\dots,\sigma_m;y)=
\det\left(\frac
    1{N\rho(u)}
    \mathcal{K}(u+\frac{2t_Nr+\sigma_i}{N\rho(u)},
u+\frac{2t_Nr+\sigma_j}{N\rho(u)};y)\right)_{i,j=1}^m\notag
\end{equation}
it follows from (\ref{4.9}), the multilinearity of the determinant and
Hadamard's inequality that
\begin{equation}
|\mathcal{R}_m^{(N)}(\sigma;y)-\mathcal{R}_m(\sigma)|\le
 m(1+\omega_N)^{m-1}\omega_N m^{m/2},\notag
\end{equation}
and hence $\epsilon_m^{(N)}\le m(1+\omega_N)^{m-1}\omega_N
m^{m/2}$. Now, by proposition \ref{prop4.1}, Stirling's formula and
the fact that $\omega_N\to 0$, 
\begin{align}\label{4.10}
&\left|\int_{\mathbb
    R^N}S_N(s,x)\rho_N(x;y(H))d^Nx-\int_0^sH''(u)du\right|\\
&\le\sum_{m=N+1}^\infty\frac{s^m}{m!}+\omega_N\sum_{m=2}^N
    \frac{s^m}{(m-1)!} (1+\omega_N)^{m-1}m^{(m+2)/2}=o(1)\notag
\end{align}
as $N\to\infty$, for a.a. $H$ $[\tilde{P}^{(N)}]$. If we combine
(\ref{4.8}) and (\ref{4.10}) we see that the theorem is proved.
\end{Proof}


\begin{thebibliography}{99}

\itemsep=\smallskipamount

\bibitem{An} C. Andr\'eief, {\em Note sur une relation les integrales
    d\'efinies des produits de fonctions,} M\'em. de la Soc. Bordeaux,
    {\bf 2} (1883), 1 - 14

\bibitem{Ba} Z. D. Bai, {\em Methodologies in spectral analysis of
    large dimensional random matrices: a review,} Statistica Sinica,
    {\bf 9} (1999), 611 - 661

\bibitem{BI} P. Bleher, A. Its, {\em Semiclassical asymptotics of 
orthogonal polynomials, Riemann-Hilbert problem, and universality 
in the matrix model,} Ann. Math., {\bf 150} (1999), 185 - 266

\bibitem{BH1} E. Br\'ezin, S. Hikami, {\em Correlations of nearby levels
induced by a random potential,} Nucl. Phys. B, {\bf 479}, (1996),
 697 - 706

\bibitem{BH2} E. Br\'ezin, S. Hikami, {\em Spectral form factor in a
random matrix theory,} Phys. Rev. E, {\bf 55}, (1997), 4067 - 4083

\bibitem{BH3} E. Br\'ezin, S. Hikami, S., {\em An extension of level-spacing
universality ,} cond-mat/9702213

\bibitem{DKMVZ} P. Deift, T. Kriecherbauer, K. T-R McLaughlin, 
S. Venakides, X. Zhou, {\em Uniform asymptotics for polynomials
  orthogonal with respect to varying exponential weights and
  applications to universality questions in random matrix theory,}
Comm. Pure. Appl. Math., {\bf 52} (1999), 1335 - 1425

\bibitem{Dy} F. J. Dyson, {\em A Brownian-motion Model for the
    Eigenvalues of a Random Matrix,} J. Math. Phys., {\bf 3} (1962),
    1191 - 1198

\bibitem{Gr} D. J. Grabiner, {\em Brownian motion in a Weyl chamber,
    non-colliding particles and random matrices,}
    Ann. Inst. H. Poincar\'e, {\bf 35} (1999), 177 - 204

\bibitem{GZ} A. Guionnet, O. Zeitouni, {\em Concentration of the
    spectral measure for large matrices,}, preprint (2000)

\bibitem{Ha} Harish-Chandra, {\em Differential operators on a semisimple Lie
algebra,} Am J. Math., {\bf 79}, (1957),  87 - 120

\bibitem{HW} D. G. Hobson, W. Werner, {\em Non-colliding Brownian
    motions on the circle,} Bull. London Math. Soc., {\bf 28} (1996),
    543 - 650

\bibitem{IZ} C. Itzykson, J. -B. Zuber, {\em The planar approximation II,} 
J. Math. Phys., {\bf 21}, (1980),  411 - 421

\bibitem{Jo1} K. Johansson, {\em Random growth and Random matrices,}
  to appear in the Proceedings of the third European Congress of
  Mathematics

\bibitem{Jo2} K. Johansson, {\em Non-intersecting paths, random
    tilings and random matrices,} in preparation

\bibitem{KM} S. Karlin, G. McGregor, {\em Coincidence probabilities,}
  Pacific J. Math, {\bf 9} (1959), 1141 - 1164

\bibitem{KS} N. M. Katz, P. Sarnak, {\em Random Matrices, Frobenius
    Eigenvalues and Monodromy,} AMS Colloquium Publications, Vol. 45, 1999

\bibitem{Kh} A. Khorunzhy, {\em On smoothed density of states for Wigner
random matrices,} Random Oper. and Stoch. Equ., {\bf 5}, (1997),
 147 - 162

\bibitem{KKP} A. Khorunzhy, B. A. Khoruzhenko, L. A. Pastur, {\em
On asymptotic properties of large rando matrices with independent
entries,} J. Math. Phys., {\bf 37}, (1996),  5033 - 5060

\bibitem{Me} M. L. Mehta, {\em Random Matrices,} 2nd ed., Academic Press, 
San Diego 1991

\bibitem{Ok} A. Okounkov, {\em Infinite wedge and measures on
    partitions,} math.RT/9907127

\bibitem{PS} L. A. Pastur, M. Shcherbina, {\em Universality of the local
eigenvalue statistics for a class of unitary invariant random matrix
ensembles,} J. Stat. Phys., {\bf 86}, (1997),  109 - 147

\bibitem{PR} E. J. Pauwels, L. G. G. Rogers, Skew-product
    decompositions of Brownian motions in {\em Geometry of Random
    Motion,} R. Durrett, M. A. Pinsky, eds., AMS Contemporary
    Mathematics, Vol. 73, 1988

\bibitem{Pi} R. G. Pinsky, {\em On the convergence of diffusion
    processes conditioned to remain in a bounded region for large time
    to limiting positive recurrent diffusion processes,} Ann. Prob.,
    {\bf 13} (1985), 363 - 378

\bibitem{Po} C. E. Porter, ed., {\em Statistical Theories of spectra:
    Fluctuations,} Academic Press, New York, 1965

\bibitem{So} A. Soshnikov, {\em Universality at the edge of the
    spectrum in Wigner random matrices,} Commun. Math. Phys., {\bf
    207} (1999), 697 - 733

\bibitem{SS1} Ya. Sinai, A.  Soshnikov, 
{\em Central limit theorem for traces of
large random symmetric matrices with independent matrix elements,}
Bol. Soc. Brasil. Mat., {\bf 29} (1998), 1 - 24

\bibitem{SS2} Ya. Sinai, A.  Soshnikov, 
{\em A refinment of Wigner's semicircle law in a neighborhood of the
  spectrum edge for symmetric matrices}, Funct. Anal. Appnl., {\bf 32}
(1998), 114 - 131

\bibitem{TW} C. A. Tracy, H. Widom, {\em Correlation Functions, Cluster
Functions, and Spacing Distributions for Random Matrices,}
J. Statist. Phys., {\bf 92}, (1998), 809 - 835

\end{thebibliography}
\end{document}